\documentclass[sigconf,authorversion]{acmart}
\usepackage{makecell}
\usepackage[ruled,vlined]{algorithm2e}
\usepackage{cleveref}
\crefname{algocf}{Algorithm}{Algorithms} 
\Crefname{algocf}{Algorithm}{Algorithms} 
\usepackage{xspace}
\usepackage{subfig}
\newcommand{\system}{\textit{Delta}\xspace}
\newcommand{\systemai}{\textit{DeltaAI}\xspace}

\AtBeginDocument{%
  }

\copyrightyear{2025}
\acmYear{2025}
\setcopyright{cc}
\setcctype{by}
\acmConference[SC '25]{The International Conference for High Performance Computing, Networking, Storage and Analysis}{November 16--21, 2025}{St Louis, MO, USA}
\acmBooktitle{The International Conference for High Performance Computing, Networking, Storage and Analysis (SC '25), November 16--21, 2025, St Louis, MO, USA}
\acmDOI{10.1145/3712285.3759821}
\acmISBN{979-8-4007-1466-5/2025/11}

\begin{document}

\title{Story of Two GPUs: Characterizing the Resilience of Hopper H100 and Ampere A100 GPUs}
\author{Shengkun Cui}
\authornote{Equal contribution.}
\affiliation{
             \institution{University of Illinois Urbana-Champaign}
             \city{Urbana}
             \country{USA}}
\email{scui8@illinois.edu}

\author{Archit Patke}
\authornotemark[1]
\affiliation{\institution{University of Illinois Urbana-Champaign}
             \city{Urbana}
             \country{USA}}
\email{apatke@illinois.edu}

\author{Hung Nguyen}
\affiliation{\institution{University of Illinois Urbana-Champaign}
             \city{Urbana}
             \country{USA}}
\email{hungnt@illinois.edu}

\author{Aditya Ranjan}
\affiliation{\institution{University of Illinois Urbana-Champaign}
             \city{Urbana}
             \country{USA}}
\email{aranjan5@illinois.edu}

\author{Ziheng Chen}
\affiliation{\institution{University of Illinois Urbana-Champaign}
             \city{Urbana}
             \country{USA}}
\email{zihengc2@illinois.edu}

\author{Phuong Cao}
\affiliation{\institution{University of Illinois Urbana-Champaign}
             \city{Urbana}
             \country{USA}}
\email{pcao3@illinois.edu}

\author{Gregory Bauer}
\affiliation{\institution{University of Illinois Urbana-Champaign}
             \city{Urbana}
             \country{USA}}
\email{gbauer@illinois.edu}

\author{Brett Bode}
\affiliation{\institution{University of Illinois Urbana-Champaign}
             \city{Urbana}
             \country{USA}}
\email{brett@illinois.edu}

\author{Catello Di Martino}
\affiliation{\institution{Nokia Bell Labs}
             \city{Sao Paulo}
             \country{Brazil}}
\email{lelio.di_martino@nokia-bell-labs.com}

\author{Saurabh Jha}
\affiliation{\institution{IBM Research}
             \city{Yorktown Heights}
             \country{USA}}
\email{Saurabh.Jha@ibm.com}

\author{Chandra Narayanaswami}
\affiliation{\institution{IBM Research}
             \city{Yorktown Heights}
             \country{USA}}
\email{chandras@us.ibm.com}

\author{Daby Sow}
\affiliation{\institution{IBM Research}
             \city{Yorktown Heights}
             \country{USA}}
\email{sowdaby@us.ibm.com}

\author{Zbigniew T. Kalbarczyk}
\affiliation{\institution{University of Illinois Urbana-Champaign}
             \city{Urbana}
             \country{USA}}
\email{kalbarcz@illinois.edu}

\author{Ravishankar K. Iyer}
\affiliation{\institution{University of Illinois Urbana-Champaign}
             \city{Urbana}
             \country{USA}}
\email{rkiyer@illinois.edu}


\renewcommand{\shortauthors}{Cui et al.}

\begin{abstract}
This study characterizes GPU resilience in \system\footnote{\system is an HPC system operated by the National Center for Supercomputing Applications (NCSA) at the University of Illinois Urbana-Champaign.}, a large-scale AI system that consists of 1,056 A100 and H100 GPUs, with over 1,300 petaflops of peak throughput. We used 2.5 years of operational data (11.7 million GPU hours) on GPU errors. Our major findings include: 
(i) H100 GPU memory resilience is worse than A100 GPU memory, with $3.2\times$ lower per-GPU MTBE for memory errors,  (ii) The GPU memory error-recovery mechanisms on H100 GPUs are insufficient to handle the increased memory capacity, (iii) H100 GPUs demonstrate significantly improved GPU hardware resilience over A100 GPUs with respect to critical hardware components, (iv) GPU errors on both A100 and H100 GPUs frequently result in job failures due to the lack of robust recovery mechanisms at the application level, and (v) We project the impact of GPU node availability on larger-scales  and find that significant overprovisioning of 5\% is necessary to handle GPU failures.
\end{abstract}




\maketitle

\section{Introduction}
Large-scale HPC systems are important not only for scientific workloads~\cite{skinner2005understanding} but also for data analytics~\cite{alam2022survey} and machine learning (ML)~\cite{huyen2022designing}.
The main components of these systems are specialized accelerators, such as GPUs, that enable acceleration of computations, such as ML training~\cite{touvron2023llama,team2023gemini}, ML inference~\cite{li2024llm}, and simulations~\cite{phillips2005scalable,shainer2009weather}.

This paper studies the resilience of A100 GPUs and compares the result with H100 GPUs of the GH200 Grace Hopper Superchips\footnote{GH200 Superchip, hereafter.}, together with their associated memory: 40 GB HBM2e on each A100 GPU, and 96 GB HBM3 on each H100 GPU, respectively.
The A100 GPU nodes and H100 GPUs (GH200 Superchip) nodes are operated as two independent systems sharing a storage cluster running the Lustre file system, allowing us to study and compare the two systems (refer to as \system).
The workflow on \system\cite{delta_2025} involves users from universities nationwide and presents a spectrum of HPC and ML workloads. 
The study uses 2.5 years of data on critical GPU errors collected across the stated GPUs, encompassing 9.6 million GPU hours of A100 GPUs and 2.1 million GPU hours of H100, a combined 11.7 million GPU hours.

\begin{figure}[!t]
    \centering
    \includegraphics[width=0.47\textwidth]{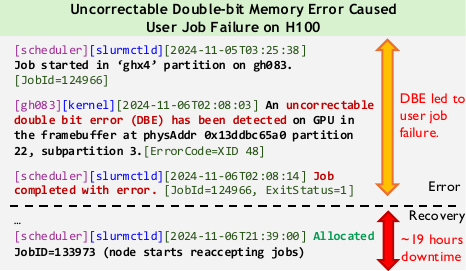}
    \caption{\small{A double-bit memory error (XID 48) occurred, and it is uncorrectable by the SECDED ECC HBM3 memory.
    Due to this double-bit error, the user job scheduled on that GPU failed, as reflected in the scheduler logs. 
    Subsequently, this uncorrectable memory error requires a node draining and reset to complete the row remapping recovery action.
    The total recovery process for this incident took 19 hours, during which the node was unavailable for accepting new jobs.
    This incident shows that a GPU error can lead to user job failures and significantly impact node availability.}}
    \label{fig:intro_example}
    \Description{A code snippet showing GSP Error Caused App Failure.}
\end{figure}

This study assesses (i) the resilience of GPU hardware and memory components; (ii) the error propagation paths in GPU memory, GPU hardware, and NVLink interconnect; and (iii) the impact of the observed GPU errors on user jobs.
\Cref{fig:intro_example} shows an example error propagation path for an uncorrectable double-bit memory error in H100 GPU that caused user job failure.
The complete recovery process required node draining and GPU reset, which took 19 hours following the error detection.

\noindent \textbf{Our major findings include:}
 

(i) H100 shows $3.2\times$ lower per-GPU mean time between errors (MTBE) compared to A100 for uncorrectable ECC memory errors. 
The per-GB MTBE of the H100's HBM3 memory is 24\% lower ($\sim\!8.5$M hours) than the A100's HBM2e memory ($\sim\!11.3$M hours).
We conjecture that the reduction in memory resilience stems from H100's higher memory capacity.

(ii) The GPU memory error-recovery mechanisms on A100 and H100 GPUs (e.g., memory row remapping, error containment)~\cite{nv_mem_err} improve GPU memory resilience and reduce service interruption. 
We observed that these mechanisms mitigate (e.g., using memory row remapping) 92\% of uncorrectable ECC memory errors on H100 GPUs. 
However, the memory error-recovery mechanism is insufficient to handle the increase in memory capacity and the corresponding increase in row remapping events on the H100 GPUs.

(iii) H100 GPUs demonstrate significantly improved hardware resilience over A100 GPUs, with respect to critical components such as GSP\footnote{A GPU system processor (GSP) is an onboard co-processor that offloads driver tasks from the CPU for latency and performance improvement.}, NVLink, and PMU SPI\footnote{PMU on an NVIDIA's GPU regulates the frequency, voltage, and power of the GPU based on various factors such as temperature and power cap. 
SPI stands for serial peripheral interface, which serves as the communication channel between peripheral hardware.}, which were major sources of job failures in A100 systems.
We attribute this to driver-level enhancements and tighter integration~\cite{nv_driver_release_notes,nvidia_grace_hopper}, which contribute to improved resilience.
Specifically, comparing H100 and A100 GPU hardware, we observed (a) a significant reduction of GSP errors on H100 (only 3 cases in our measurement period) and (b) the elimination of PMU SPI error propagations, which on A100 GPUs can lead to MMU errors 88\% of times with 90\% leading to user job failures, (c) no NVLink errors on H100 GPUs during the measurement period.

(iv) GPU errors on both A100 and H100 GPUs frequently result in job failures due to the lack of robust recovery mechanisms at the application level. 
Except for MMU and NVLink errors, other GPU errors cannot be handled by application-level mechanisms, resulting in close to $100\%$ job failure rate.
The underlying cause of job failures differs by GPU type: hardware errors are the predominant cause in A100 GPUs, whereas memory errors are the primary cause in H100 GPUs.
Overall, we find that application-based recovery strategies are largely ineffective; hence, there is a compelling need to improve resilience at the GPU memory and hardware level.

(v) The overall availability per-GPU node is approximately $99.4\%$ for A100 GPUs and $99.3\%$ for H100 GPUs, corresponding to a downtime between 9--10 minutes per day.
We projected the impact of this measured availability on larger scales via emulation.
For example, to maintain 99.9\% availability at the job level, overprovisioning of 5\% would be necessary.
While at first glance, such overprovisioning would appear to be a small cost, for the above example, it would cost over \$1 million per month.
If GPU node availability were improved to 99.9\%, the required overprovisioning would reduce by $2.5\times$.

\textbf{Putting the paper in perspective.} Previous studies on characterizing GPU resilience in large-scale systems~\cite{debardeleben2014gpu,di2014lessons,tiwari2015understanding,tiwari2015reliability,gupta2015understanding,gupta2017failures,ostrouchov2020gpu,nie2016large,nie2017characterizing,oles2024understanding} focus on GPU memory errors in older GPU generations (Tesla, Kepler, and Volta) that lack the latest resilience (e.g., row remapping, error containment, NVLink CRC-retry) and performance features (e.g., GPU System Processor) introduced in NVIDIA Ampere-generation GPUs. 
A recent study from Meta~\cite{kokolis2024revisiting} characterizes cluster-level resilience for two large-scale machine learning clusters equipped with A100 GPUs.
Our paper provides a deeper understanding of GPU errors and failures of two recent generations of GPUs, A100 and H100, and their impact on a broad set of HPC/ML applications. 
To the best of our knowledge, this is the first study on A100 and H100 GPU errors in HPC/ML systems.

\textit{All software, datasets, and analysis scripts are available in the AD/AE materials and at https://doi.org/10.5281/zenodo.15287639.}

\section{Background}
This section provides information on (i) \system specification, (ii) \system overall GPU utilization, GPU operational environment, and workloads, (iii) critical GPU error categories used in this study, and (iv) GPU error management and recovery.

\begin{figure}[t]
    \centering
    \includegraphics[width=0.42\textwidth]{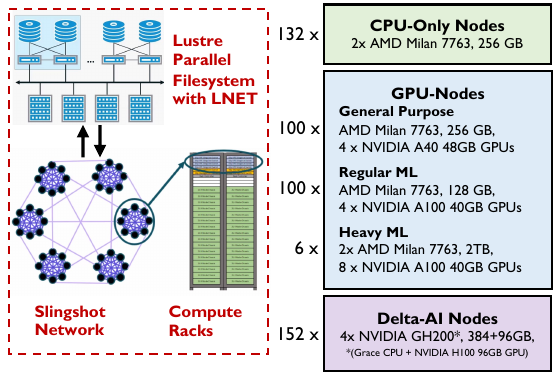}
    \caption{\small{System architecture and specifications of \system. This study focuses on the H100 and A100 GPU nodes.}}
    \label{fig:delta_arc}
    \Description{Delta System Diagram.}
\end{figure}
\subsection{\system Specifications}
\textbf{\system Specifications.} 
\Cref{fig:delta_arc} shows the layout of \system, which consists of a cluster of 106 4- and 8-way A100 40 GB GPU nodes with 448 A100 GPUs, and in addition, it has a second cluster of 152 4-way GH200 nodes with 608 H100 96 GB GPUs, a total of 1,056 A100 and H100 GPUs. The two clusters share a common storage cluster running Lustre Filesystem.
Our study  focuses on the resilience of A100 GPUs and H100 GPUs (the GPU of GH200), as they are optimized for AI/ML workloads, exhibit the highest utilization, and incorporate the latest resilience features.

Note that the H100 GPUs studied here are integrated in GH200 Superchips, tightly coupled to NVIDIA Grace CPUs via NVLink-C2C interconnect, distinguishing them from discrete H100 GPUs. 
While the H100 microarchitecture and memory specifications are consistent across both variants, differences in CPU–GPU integration may lead to variations in resilience characteristics.

\subsection{\system GPU Operational Environment}
A100 and H100 GPUs ran in comparable operational conditions in terms of (i) utilization, (ii) cooling/temperature, and (iii) workloads.

\textbf{Overall Utilization.} \system's NVIDIA A100 GPUs are frequently scheduled and utilized, with an average GPU utilization of 51\% during the operational period.
NVIDIA H100 GPUs show a slightly lower average utilization (41\%) than A100 GPUs.

\textbf{Cooling and Temperature.} \system's A100 and H100 GPUs are liquid-cooled from shared facility water supply through independent cooling loops. At average utilizations, their mean temperatures are 40$^\circ$C (A100) and 37$^\circ$C (H100), with maximums of 49$^\circ$C and 48$^\circ$C, respectively, indicating comparable operational conditions.

\textbf{Workloads.}
We use the allocated \system projects' Field-of-Science distributions as a proxy for workload characterization. 
Both GPUs handled similarly diverse workloads with comparable distributions across the top five fields (A100s/H100s): Computer Science (30.4/32.8\%), AI \& Intelligent Systems (18.1/29.1\%), Applied CS (4.7/4.2\%), Biophysics (5.1/1.7\%), Materials Engineering (3.6/1.4\%) for A100/H100, respectively.

\subsection{NVIDIA GPU Error Categories}
\label{sec:gpu_errors}

NVIDIA GPU errors are reported as XID errors. 
In this study, we selected a subset of XID errors that are described as common and high-impact by NVIDIA's Developer Manuals~\cite{nv_xid_doc,nv_mem_err}, NVIDIA Developer Forums and Blogs, and \system site reliability engineers (SREs). 
We primarily collected errors and their associated recovery events.
The selected XID errors/events indicate GPU issues that often cannot be resolved without SRE's interventions (e.g., node service and GPU replacement).
The selected XIDs and their corresponding GPU errors are described in~\Cref{sec:res}, ~\Cref{tab:gpu_res_stats}.
We categorize the selected GPU errors into three categories: (i) GPU hardware, which includes all onboard hardware except for GPU memory and NVLink interconnect, (ii) NVLink interconnect, and (iii) GPU memory.  
Note that General GPU Software Error (XID 13) and Reset Channel Verification Error (XID 43) 
are usually caused by user jobs and do not impact the health of the GPU~\cite{nv_xid_doc}; we excluded those errors from our study.

\textbf{GPU Hardware Errors.} 
The critical GPU hardware errors we studied include MMU\footnote{The memory management unit (MMU) provides essential memory I/O functionalities.} errors, GPU Fallen Off the Bus errors, GSP  errors, and PMU communication errors. 
We do not consider other GPU hardware errors in our study.  
GPU hardware errors can lead to user job failures, GPU halt, and data corruption.  
Among those errors, GPU Fallen Off the Bus and GSP errors lead to GPU failures, and manual GPU resets or node reboots are required to recover from the error~\cite{nv_xid_doc}.
\system SREs monitor GSP errors closely to ensure timely recovery to maintain GPU availability.

\textbf{GPU Interconnect (NVLink) Errors.} GPU-GPU NVLink errors are caused by faulty GPU hardware, connectors, or improper connector installation during system integration, and can lead to GPU unavailability and user job failures. NVLink errors impede data transfer between GPUs and reduce computational throughput.
A GPU reset or node reboot is required to clear NVLink errors~\cite{nv_xid_doc}.

\textbf{GPU Memory Errors.}
GPU memory errors included in this study are double-bit errors (DBEs) and consecutive single-bit errors (SBEs)\footnote{Consecutive SBEs are multiple single-bit error (SBE) occurrences at the same memory location.}. 
Individual SBEs are not logged, as they are automatically corrected by ECC.
DBEs and consecutive SBEs are considered uncorrectable ECC memory errors by the NVIDIA driver, and they trigger downstream error-recovery mechanisms~\cite{nv_xid_doc,nv_mem_err}, which are introduced in~\Cref{sec:gpu_fault_mech}. 
Failures in these mechanisms can lead to GPU/node failures and require GPU or node reboots to recover~\cite{nv_xid_doc}.  
\system SREs continuously monitor uncorrectable ECC memory errors and error-recovery failures to ensure timely replacement of faulty GPUs. 

\label{sec:gpu_fault_mech}
\begin{figure}[t]
    \centering
    \includegraphics[width=0.43\textwidth]{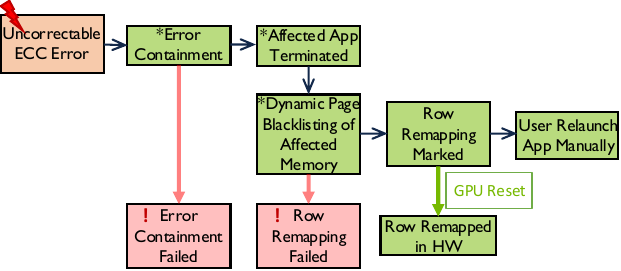}
    \caption{\small{NVIDIA memory error recovery process for A100 and H100 GPUs.}}
    \label{fig:gpu_recovery_path}
    \Description{A plot showing NVIDIA GPU recovery path}
\end{figure}

\subsection{NVIDIA GPU Error Management}
Here, we provide an overview of the resilience architecture of NVIDIA A100 and H100 GPUs.

\textbf{GPU Memory.} 
\Cref{fig:gpu_recovery_path}  
shows the uncorrectable ECC memory error-recovery process~\cite{nv_mem_err} for A100 and H100 in more detail.
The primary mechanism for mitigating uncorrectable ECC memory errors for A100 and H100 GPUs is \textit{row-remapping}, wherein the faulty memory row is replaced with a spare row, and a  
row-remapping event (RRE) is logged. 
The actual row remapping happens at the next GPU reset (e.g., during node reboot or maintenance).
If there are no spare memory rows, a row remapping failure (RRF) is indicated ~\cite{nv_xid_doc,nv_mem_err}.

A100 and H100 GPUs support online recovery mechanisms such as \textit{error containment} and \textit{dynamic page offlining}~\cite{nv_xid_doc,nv_mem_err} for mitigating uncorrectable ECC memory errors with minimal node interruption. 
The dynamic page offlining marks the faulty memory page as unusable without requiring a GPU reset to maintain availability. 
The error containment procedure terminates user processes using the faulty memory address to prevent error propagation to other applications. 
Successful error containment is logged as a Contained Memory Error, whereas an unsuccessful error containment is logged as an Uncontained Memory Error.
Failure in a row-remapping or error containment can cause a GPU failure that requires a GPU reset or node reboot.
\system SREs monitor row-remapping failures and replace GPUs that repeatedly emit such errors. 

\textbf{GPU Hardware.} 
While the GPU caches and memory are SECDED protected,  information on failure-recovery mechanisms on GPU hardware, including peripheral hardware such as GSP, PMU, or SPI communication channels, is limited.

\textbf{GPU Interconnect (NVLink).} 
NVLink employs Cyclic Redundancy Checks (CRCs) for error detection to ensure the integrity of flow control digits and data.
Upon encountering a CRC checksum error, NVLink retries packet transmissions from the last-known good packet.
\section{Methodology}
\label{sec:data_method}

\subsection{Data Sources}
\label{sec:data}

\begin{figure}[!t]
    \centering
    \includegraphics[width=0.43\textwidth]{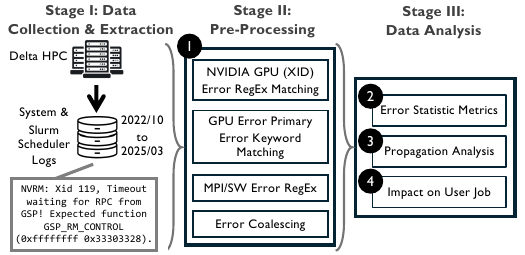}
    \caption{Overview of our data collection, processing, and analysis pipeline.}
    \label{fig:pipeline}
    \Description{Pipeline figure showing data collection, processing and analysis.}
\end{figure}

Our analysis was performed on data collected from~\system over its operational period: (i) 895 days from October 2022 to March 2025 for A100 GPUs and (ii) 146 days from October 2024 to March 2025 for H100 GPUs, covering 11.7 million GPU hours.
This section describes data sources for \textit{Stage I}: data collection and extraction in the pipeline in~\Cref{fig:pipeline}. 

\textbf{System logs.} 
System logs collected from all compute nodes capture events across system components.
We created a set of regular expression (RegEX) patterns and used it to extract GPU error-recovery log entries by referring to NVIDIA XID messages~\cite{nv_xid_doc} from the system logs (\Cref{fig:pipeline}, item (1)). 
The GPU error logs were our major sources of error and recovery information. 

\textbf{Slurm scheduler database.} \system uses the Slurm Workload Manager~\cite{slurm_log} (``Slurm scheduler'' hereafter) for scheduling user jobs. 
The Slurm scheduler database keeps track of user job scheduling events, including the start and end times, the scheduled nodes, resource usage, job status, exit codes, and the srun command line. 
We used the Slurm database for user job failure characterization.

\textbf{NVIDIA DCGM database.} 
\system uses NVIDIA Data Center GPU Manager (DCGM) to collect metrics and status data from all GPUs with a one-minute granularity. We used DCGM metric data to characterize GPU utilization.

\subsection{Data Processing Pipeline}
\label{sec:method}

This section focuses on \textit{Stage II and III} of the data processing pipeline in ~\Cref{fig:pipeline}, which pre-processes the raw logs, compute error counts, mean times between errors (MTBE), error propagation, and impact on user jobs.

\begin{algorithm}[t!]
\footnotesize
\DontPrintSemicolon 
\SetKwInOut{Inputs}{Input}
\SetKwInOut{Output}{Output}
\SetAlgoNlRelativeSize{-1}
\caption{Error Coalescing and Persistence Analysis.}\label{alg:error_persistence}
\Inputs{Error logs with timestamps $E = \{(e_1, t_1), \dots, (e_n, t_n)\}$, regex patterns $\mathcal{R} = \{r_1, r_2, \dots\}$, time window $\Delta t$}
\Output{Coalesced errors with persistence duration $E'$}
$E' \gets \varnothing$ // Initialize output set

\ForEach{pattern $r \in \mathcal{R}$}{ 
    
    $E_r \gets \{(e_i, t_i) \in E \mid e_i \text{ matches } r\}$ Filter errors with regex
     $i \gets 1$ 
    
    // Loop through errors in a matched group
    
    \While{$i \leq |E_r|$} { 
        $(e_{\text{first}}, t_{\text{start}}), t_{\text{latest}} \gets (e_i, t_i), t_i$\; 
        // Loop through later errors within the matched group 
        \While{$i+1 \leq |E_r|$} {
            $(e_{\text{next}}, t_{\text{next}}) \gets (e_{i+1}, t_{i+1})$\;
            // Error has identical message and is close in time
            \If{$e_{\text{next}} = e_{\text{first}}$ \textbf{and} $t_{\text{next}} - t_{\text{latest}} \leq \Delta t$}{
                $t_{\text{latest}} \gets t_{\text{next}}$ //Discard latest error $i \gets i + 1$\; 
            }
            \Else{\textbf{break}}
        }
        // Store coalesced error and persistence duration
        
        Add $(e_{\text{first}}, t_{\text{start}}, t_{\text{latest}} - t_{\text{start}})$ to $E'$
        
        $i \gets i + 1$ // Move to the next unprocessed error
    }
}
\Return{$E'$}\;
\end{algorithm}

\textbf{Error Coalescing Analysis.\label{met:error_col}} The error coalescing step in~\Cref{fig:pipeline}, item (1) filters out duplicated errors. While most errors are logged as isolated events, there are frequent periods during which the same error is logged repeatedly in close succession, i.e., there are error bursts. 
During these bursts, the system continually detects and attempts to recover 
from errors, which could lead to either system recovery or failure.
To prevent over-counting, we assume that identical error logs within a short time interval ($\Delta t$) from the same GPU are caused by the same issue.
Thus, the error coalescing step counts only the first occurrence by combining identical error log lines from the same GPU within a predefined time interval ($\Delta t$) into a single error (see Algorithm~\ref{alg:error_persistence}).
The remaining analyses in this paper were conducted on errors after the coalescing.
We set $\Delta t=5$ as decreases in $\Delta$ results in more duplicated logs, while further increase in $\Delta$ results in negligible changes in coalesced log count.

\textbf{Error Statistic Metrics.} 
Using the coalesced error logs as input, the pipeline computes standard error statistics metrics including the error count and the MTBE as in~\cite{oles2024understanding} (\Cref{fig:pipeline}, item (2)). MTBE allows fair comparison between GPU types with different total operational hours.
For errors that are not directly countable from XID logs, we estimate their occurrence by correlating related XIDs within the same time interval~\cite{oles2024understanding,nv_xid_doc,nv_mem_err}. 
In particular, the number of uncorrectable ECC memory errors can be inferred by summing up the number of RREs and RRFs, as they are mutually exclusive outcomes of an uncorrectable ECC memory error recovery event.
Subsequently, the number of consecutive SBEs can be obtained by subtracting the number of DBEs from the number of uncorrectable ECC memory errors.

We additionally computed system-wide MTBE and derived per-node MTBE by normalizing the error count using the number of GPU nodes in \system. 
The per-node MTBE indicates the operational hours a single \system GPU node can function before encountering an error.
For GPU memory resilience characterization, we additionally derived per-GPU MTBE by normalizing the error count using the number of GPU in a node and per-GB MTBE by normalizing the error count with per GPU memory capacity in GB.
The per-GPU MTBE reflects the operational hours a single GPU can function before encountering an error.

\textbf{Error Propagation Analysis.} We performed error propagation analysis (\Cref{fig:pipeline}, item (3)) to capture how errors propagate within a GPU and across different GPUs while measuring the propagation time. 
The propagation probability from GPU error \( e_1 \) to \( e_2 \) is defined as  
$P(e_2|e_1) = \frac{\# e_2}{\text{Total } \# e_1}, \quad t_{e_2} - t_{e_1} \leq \Delta t.$
A propagation path is created if \( e_2 \) occurs immediately after \( e_1 \) within a predefined time window \(\Delta t\). If there is no succeeding error $e_2$ after $e_1$ within $\Delta t$, then $e_1$ is a terminal error that does not propagate.
For intra-GPU\footnote{GPU devices are identified by their node ID and PCI Express bus address.} propagation, we require errors $e_1$ and $e_2$ to be on the same GPU device, whereas for the inter-GPU propagation, $e_1$ and $e_2$ are from two distinct GPUs on the same node. 
We recorded the time difference between the initial ($e_1$) and subsequent errors ($e_2$), referred to as the \textit{propagation time}, for each propagation event. 
A shorter propagation time suggests a higher correlation between $e_1$ and $e_2$.
We applied the same $\Delta t$ selection criteria as the error coalescing analysis and selected the $\Delta t=5$ seconds in error propagation analysis.

\textbf{User Job Impact Analysis.}
\label{sec:app_impact_ana}
The user job impact analysis step (\Cref{fig:pipeline}, item (4)) associates GPU errors with failed user jobs to characterize the impact of GPU errors on user jobs.~\Cref{sec:app} provides detail on this analysis.

\begin{table*}[ht]
\centering
\caption{\label{tab:gpu_res_stats}\system NVIDIA Ampere A100 and Hopper H100 (GPU of GH200 Superchip) GPU resilience statistics.}
\resizebox{0.97\linewidth}{!}{
\begin{tabular}{
>{\centering\arraybackslash}m{1.2cm}|
>{\centering\arraybackslash}m{1.7cm}|
>{\centering\arraybackslash}m{1.5cm}|
>{\centering\arraybackslash}m{5cm}|
>{\centering\arraybackslash}m{3.5cm}|
>{\centering\arraybackslash}m{1.0cm}|
>{\centering\arraybackslash}m{1.0cm}|
>{\centering\arraybackslash}m{1.0cm}|
>{\centering\arraybackslash}m{1.0cm}|
>{\centering\arraybackslash}m{1.0cm}|
>{\centering\arraybackslash}m{1.0cm}
}

\toprule
\textbf{Event} & \textbf{Abbr.} & \textbf{Category} & \textbf{Description} & \textbf{Recovery Action} & \multicolumn{2}{c|}{\textbf{Count}} & \multicolumn{4}{c}{\textbf{MTBE (hrs)}} \\
\textbf{Code} & & & & & \textbf{A100} & \textbf{H100} & \multicolumn{2}{c|}{\textbf{System-wide}} & \multicolumn{2}{c}{\textbf{Per Node}} \\
& & & & & & & \textbf{A100} & \textbf{H100} & \textbf{A100} & \textbf{H100} \\
\midrule

XID 31 & MMU Error & Hardware & GPU memory management unit (MMU) error.  
& MMU error due to invalid memory access or driver/hardware bugs.  
& 8,863 & 1,737 & 2.4 & 2 & 257 & 307 \\

\hline
XID 48 & DBE & Memory & Double bit ECC memory error (DBE).  
& Triggers RRE; GPU reset or node reboot is needed to clear this error. 
& 1 & 17 & -- & 206 & -- & 31,330 \\

\hline
-- & Consecutive SBEs & Memory & Consecutive single-bit ECC memory errors (SBEs).  
& Triggers RRE; GPU reset or node reboot is needed to clear this error. 
& 33 & 7 & 651 & 501 & 68,996 & 76,087 \\

\hline
-- & Uncorrectable ECC memory Errors & Memory & Consecutive SBEs or a DBE.  
& Triggers RRE; GPU reset or node reboot is needed to clear this error. 
& 34 & 24 & 632 & 146 & 66,967 & 22,192 \\

\hline
XID 63 & RRE & Memory & Row remapping event triggered by uncorrectable MBE: one DBE or two SBEs at the same memory address.  
& GPU reset needed for row remapping. 
& 34 & 16 & 632 & 219 & 66,967 & 33,288 \\

\hline
XID 64 & RRF & Memory & Row remapping failure of a row remapping event.  
& GPU reset is needed to clear this error. 
& 0 & 8 & -- & 438 & -- & 66,576 \\

\hline
XID 74 & NVLink Error & Inter-connect & NVLink error indicating connection issues between GPUs via NVLink interconnection.  
& GPU reset or SRE intervention required. 
& 1,922 & -- & 11 & -- & 1,185 & -- \\

\hline
XID 79 & GPU Fallen Off the Bus Error & Hardware & GPU has fallen off the system bus and is not reachable, typically because of driver or hardware issues.  
& GPU reset or SRE intervention required. 
& 10 & -- & 2,148 & -- & 227,668 & -- \\

\hline
XID 94 & Contained Memory Error & Memory & Uncorrectable contained ECC error, indicating successful containment.  
& Not specified. 
& 13 & 14 & 1,652 & 250 & 175,144 & 38,043 \\

\hline
XID 95 & Uncontained Memory Error & Memory & Uncontained memory error, indicating failure in containment.  
& GPU reset or SRE intervention required. 
& 11 & 19 & 1,953 & 184 & 206,989 & 28,032 \\

\hline
XID 119/120 & GSP Error & Hardware & NVIDIA GPU Systems Processor (GSP) error.  
& GPU reset or SRE intervention required. 
& 3,857 & 3 & 6 & 1,168 & 590 & 177,536 \\

\hline
XID 122/123 & PMU SPI Error & Hardware & PMU SPI read/write failure, indicating failed communication with the PMU.  
& Not specified. 
& 77 & -- & 279.0 & -- & 29,570 & -- \\
\hline 
\multicolumn{11}{l}{\small *NVIDIA A100/H100 GPU supports page retirement and up to 512 row remappings (RRE); previous generations support only 64 page retirements (no row remapping support).}\\
\multicolumn{11}{l}{\small *Row remapping, contained memory error, and uncontained memory error are new resilience features introduced starting with the NVIDIA Ampere for managing uncorrectable memory errors.} \\
\multicolumn{11}{l}{\small *Per-node MTBE (hrs) is derived by multiplying system MTBE by the number of GPU nodes of that GPU type. The number of nodes and GPUs per node type are specified in~\Cref{fig:delta_arc}.} \\
\multicolumn{11}{l}{\small *All XID events presented, except for row remapping events, are errors. However, for simplicity, we treat all XID events as errors in this paper.} \\
\multicolumn{11}{l}{\small *Uncorrectable ECC and Consecutive SBEs are inferred from the corresponding recovery event RRE and RRF (see~\Cref{sec:method}).} \\
\bottomrule 
\end{tabular}  
}
\end{table*}

\section{Characterizing GPU Resilience}
\label{sec:res}
This section characterizes and compares the resilience of \system's NVIDIA A100 (Ampere) and H100 (Hopper) GPUs. 
Specifically, we discuss  error statistics and error propagation of GPU errors in three categories: (i) GPU memory, (ii) GPU hardware, and (iii) NVLink interconnect, as described in~\Cref{sec:gpu_errors} and~\Cref{tab:gpu_res_stats}.
These errors are critical because they interrupt user jobs and lead to unplanned node downtime, as we show in~\Cref{sec:app}. We first highlight error statistics and key findings from our analysis and then focus on error propagation for each of the three error categories on \system's workload.
Note that we do not directly compare \system with \textit{Blue Waters}\cite{di2014lessons}, \textit{Titan}\cite{nie2016large}, or \textit{Summit}~\cite{oles2024understanding} as those systems used older GPUs lacking latest performance and resilience features central to our study, e.g., GSP, dynamic page offlining, row remapping, memory error containment, and NVLink recovery.

\subsection{Error Statistics and Result Highlights} 
\label{sec:res_hl}
\Cref{tab:gpu_res_stats} summarizes the selected critical GPU error statistics for A100 GPUs and H100 GPUs during the operational period, including error count, system-wide mean time between errors (MTBE) and per-node MTBE.
As described in~\Cref{sec:gpu_errors}, these selected errors are critical XID errors that can lead to user job interruption and node downtime, requiring manual SRE intervention for recovery.
In the operational period, a total of 14,821 critical errors (listed in~\Cref{tab:gpu_res_stats}) were recorded for \system's A100 GPUs, with a system-wide MTBE of 1.4 hours and node MTBE of 154 hours. 
\system's H100 GPUs encountered 1,821 GPU errors, with a system-wide MTBE of 1.9 hours and node MTBE of 292 hours, higher than A100 nodes.
Next, we use the results from~\Cref{tab:gpu_res_stats} to assess the resilience characteristics of GPU memory, GPU hardware, and NVLink interconnect in greater detail. 

\textbf{GPU Memory.}
Recall that, each A100 GPU incorporates 40 GB of HBM2e memory, and each H100 GPU incorporates 96 GB of HBM3 memory.
H100’s shows $3.2\times$ lower per GPU MTBE for uncorrectable ECC memory errors compared to A100’s, despite comparable per-GB MTBE.
This reduced resilience likely stems from H100’s higher memory density, which trades resilience for capacity and performance. 
The observed lower per node MTBE of RREs and higher RRF counts on H100 GPUs further suggest that current resilience features are insufficient for increased memory capacity. 
As a uncorrectable memory error can cause a multi-GPU job to fail, enabling uninterrupted execution in such cases at scale remains an open challenge. We present the detailed analysis below.

(i) \system's H100 GPUs exhibit lower memory resilience than
the A100 GPUs. Specifically, the per GPU MTBE is 3.2$\times$ lower on H100 GPUs at 88,768 hours versus 283,271 hours for A100 GPUs.
Based on the per GPU MTBE, we calculated the per GB MTBE to be 8,521,728 hours\footnote{We calculated the per GB MTBE by multiplying the per node MTBE with the total memory of all GPUs in GB for that node. 
For example, in a 4-way H100 GPU node with 96 GB of memory, the per GB MTBE = $22,192\times 96 \times 4 = 8,521,728$ hours.} for H100's HBM3 memory vs. 11,330,826 hours for A100's HBM2e memory, a 24\% reduction.

We attribute the decrease in resilience is primarily due
to the higher memory capacity (96 GB vs. 40 GB, a $2.4\times$ increase),
which increases the chances of bit flips. We additionally
hypothesize that H100 memory resilience is worse due to (a) a lower
signaling voltage that increases susceptibility to bit flips~\cite{wu_atc_hbm} and
(b) an increased number of stacks that make heat dissipation challenging and degrade the resilience of memory modules, of the HBM3 memory.

(ii) Uncorrectable ECC memory error-recovery mechanisms  (e.g., row remapping and error containment, see~\Cref{sec:gpu_fault_mech}) improve GPU memory resilience and reduce service interruptions~\cite{nv_mem_err}. The error recovery mechanisms mitigate uncorrectable memory errors with a probability of 0.92 on H100 according to error propagation analysis in~\Cref{sec:error_prop}.
However, these mechanisms may not scale well with the increase in GPU memory capacity in H100.
For example, the available spare rows for row remapping are capped at the same 512 rows~\cite{nv_mem_err}, which is not proportional to the $2.4\times$ increase in memory capacity. 
Such insufficiency can be evident from the significantly lower per node MTBE of RRE on the H100 GPUs than the A100 GPUs, indicating more frequent recovery events.
In addition, we observed 8 RRFs on H100 GPUs during the early operational period, which indicates memory recovery failure due to exhaustion of spared memory rows. We have yet to observe an RRF on the A100 GPUs during the much longer operational period.

\textbf{GPU Hardware and NVLink Interconnect.}
\label{sec:hardware_error}
GPU hardware such as GSP, PMU SPI, and NVLink are critical components from a resilience perspective for A100 GPUs, leading to node downtime and job failures.  
In this context, H100 GPUs demonstrate significant improvements in GPU hardware resilience, especially in critical components such as the GSP, PMU SPI, and NVLink.
We believe that these improvements are likely due to the tightly integrated heterogeneous CPU-GPU architecture of GH200~\cite{nvidia_grace_hopper} and driver-level enhancements~\cite{nv_driver_release_notes}.
We present a detail analysis below.

(i) Among the GPU hardware components on A100 GPUs, GSP, intended as a performance enhancement, is the most vulnerable due to its lack of robust detection and recovery.
Our error propagation analysis (\Cref{sec:error_prop}) shows that over 99\% of GSP errors put the GPU in an error state and lead to job failure if encountered.
A GSP error requires a GPU reset or a node reboot to recover, introducing significant overheads.
In addition, PMU SPI communication errors propagate downstream and cause MMU errors with a probability of 0.88, leading to MMU errors, which then, in turn, result in a job failure over 90\% of the time. 
Although PMU SPI communication errors are high-impact errors, they are not highlighted in NVIDIA's developer's manual~\cite{nv_xid_doc}.

(ii) Despite error detection mechanisms such as CRC and recovery mechanisms such as message retransmitting, NVLink GPU interconnect errors are 
frequent (1,922 total NVLink errors) on A100 GPUs, with a node MTBE of 1,185 hours (system-wide MTBE of 11 hours); 42\% (801) of the NVLink errors affected two or more GPUs. 
Although NVIDIA~\cite{nv_xid_doc} indicates that a GPU reset or a node reboot is required to clear NVLink, \system SREs suggested that NVLink errors were largely benign and did not always lead to job failures.
Indeed, the user job impact analysis shows that an NVLink error has a 54\% chance of leading to a job failure (see~\Cref{sec:app}). 
When an NVLink error is observed but does not affect a user job, it may be because NVLink is primarily used for communication rather than computation in many jobs or because multiple NVLink errors within the same job have a consolidated impact, resulting in high occurrence rates but minimal application disruption.

(iii) H100 GPUs have substantially improved GPU hardware resilience over A100 GPUs: only 3 GSP errors were observed during the measurement period.
Moreover, the number of GPU Fallen off the bus errors, NVLink errors\footnote{We tested the NVLink to ensure that the NVLinks were enabled.}, and PMU SPI errors were not observed in H100 GPUs.
We attribute this improvement to (a) the better packaging and tight of CPU and GPU into a single heterogeneous compute module, which significantly reduces integration errors and enhances the resilience of the CPU-GPU complex, and (b) the NVIDIA driver upgrades appear to have improved GSP and NVLink stability (also observed by the \system SREs).

\textbf{GPU Errors Temporal Analysis.} Temporal analysis revealed that GPU error rates followed the classic ``bathtub'' reliability curve.
In the pre-operational period, i.e., the ``infant-mortality'' phase, \system underwent extensive testing, hardware replacements, and software fixes, during which the system-wide MTBE was 0.15 hours\footnote{Evaluated using the critical-errors listed in \Cref{tab:gpu_res_stats}.}.
In the operational period, the system-wide MTBE rose $10\times$ to 1.4 hours, marking a transition to the ``normal-life phase'' with lowered error rate.
Furthermore, estimation of the hazard-rate ($H(t)=\int h(t)\,dt$) and cumulative hazard-rate using Nelson-Aalen estimator showed that GPU hazard-rate varies over time but exhibits no clear temporal trends during the operational period (e.g., hazard rate with mean 0.7, std 7.94 for A100 GPUs).

\textbf{GPU Failure and Replacement.}
During the measurement period (895 days) of A100 GPUs, four A100 GPUs were replaced due to GPU's failure to boot-up. 
In comparison, during the measurement period (146 days) of the H100 GPUs, two GPUs were replaced due to uncorrectable memory errors and row remapping failures. 
This provides an additional support for the observed result that memory is a resilience weak link on H100 GPUs, while the A100 GPUs are more susceptible to errors from GPU hardware components.

\subsection{GPU Error Propagation}
\label{sec:error_prop}

\begin{figure}[t!]
    \centering
    \includegraphics[width=0.90\linewidth]{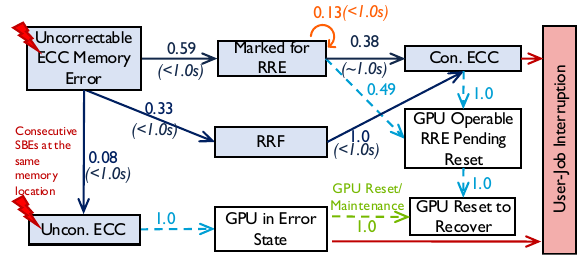}
    \caption{\label{fig:gpu_err_dbe_hopper} Intra-GPU uncorrectable memory error recovery paths in H100 GPUs. Numbers on the edges show propagation probability. The precise sub-second timing information is not available in the H100 nodes' system logs. All propagation time for H100 GPUs memory errors were within one second.}
\Description{Graph showing memory error propagation probabilities.}
\end{figure}

This section describes results on both \textit{intra-GPU} and \textit{inter-GPU error propagation} during the operational period for \system's A100 and H100 GPUs. 
Understanding GPU error propagation reveals resilience weak links between GPU components. 
We break down the error-recovery propagation into three categories for the GPU errors listed in~\Cref{tab:gpu_res_stats}: (i) GPU memory, (ii) GPU hardware, and (iii) NVLink interconnect, and
estimate the propagation probabilities from GPU error logs (see~\Cref{sec:method}).
The propagation paths in~\Cref{fig:gpu_err_dbe_hopper,fig:gpu_error_hardware,fig:gpu_nvl_prop} are highlighted with lightning signs that indicate the beginning of each path. 
If two succeeding errors occur in close successions, i.e., a short propagation time, it suggests causality.

For GPU memory errors, we observed propagation primarily on H100 GPUs, which coincides with our findings on worsened memory resilience in~\Cref{sec:hardware_error}.
For GPU hardware and NVLink interconnect errors, we observed error propagation paths primarily on A100 GPUs because of the improved hardware resilience of H100 as discussed in~\Cref{sec:hardware_error}.

\subsubsection{\textbf{GPU Memory-Error Propagation.}} 
Intra-GPU uncorrectable memory error recovery paths are shown in
~\Cref{fig:gpu_err_dbe_hopper} for H100 GPUs during the operational period. 
Memory error recovery path for A100 GPUs were a subset shown in~\Cref{fig:gpu_err_dbe_hopper}.
Hence, we only show the memory propagation path for H100 GPUs.

\textbf{Successful Error Recovery and Containment.} As shown in ~\Cref{fig:gpu_err_dbe_hopper}, row remapping recovery (RRE) triggered by an uncorrectable ECC memory error has a success rate of 0.59 on H100 GPUs.
For the 33\% of the row remapping events that fail (RRF), the GPU still contains the uncorrectable memory error because only the affected user jobs were terminated, and only the faulty page was offlined.  
Overall, considering both uncorrectable memory error recovery paths (RRE and error containment after an RRF), the impact of uncorrectable memory errors was alleviated 92\% of the time on H100, while the GPU can remained operable. 
Such uninterrupted operations were not achievable on previous-generation GPUs (e.g., Kepler in~\cite{di2014lessons,tiwari2015reliability} and Volta in~\cite{oles2024understanding}), as an uncorrectable memory error would immediately cause user job interruption and put GPU in an error state, necessitating a GPU reset to recover~\cite{nv_xid_doc}.  

\textbf{Unsuccessful Error Containment.} 
The above uncorrectable ECC memory error recovery or containment process can fail, resulting in uncontained memory errors (\Cref{sec:gpu_fault_mech}).
The error containment process failed 8\% of the times on H100 GPUs during the operational period (see~\Cref{fig:gpu_err_dbe_hopper}).
Moreover, as informed by the \system SREs, uncontained memory errors can be highly bursty and persistent and may spam the console logs, consuming useful compute cycles and leaving the GPU inoperable. 
We did not observed highly bursty and persistent uncontained memory error in A100 or H100 GPUs during the operational period.

Overall, the propagation analysis suggests that the new memory error recovery mechanisms on H100 GPUs alleviated impact of uncorrectable memory errors 92\% of the time. 
That said, \system's H100 GPUs exhibited significantly more memory error recovery propagations than A100 GPUs, highlighting the worsened memory resilience compared to A100s.
In addition, the highly bursty and persisting nature of uncontained memory errors can lead to node/system operation disruptions, as we learned from \system SREs.

\subsubsection{\textbf{GPU Hardware-Error Propagation.}}

\begin{figure}[t!]
    \centering
    \includegraphics[width=0.90\linewidth]{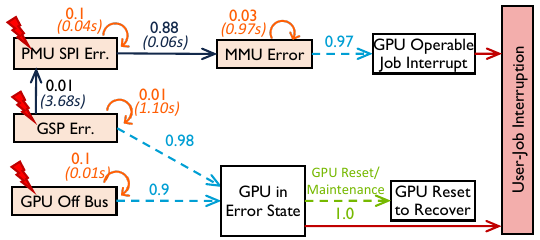}
    \caption{\label{fig:gpu_error_hardware}Intra-GPU hardware error propagation probabilities in A100 GPUs. Numbers on the edges show propagation probabilities and average propagation time in seconds.
    }
\Description{Plot showing hardware error propagation with probabilities}
\end{figure}

\label{sec:hardware_prop}
This section primarily focuses on A100 because the H100 hardware (these components) experienced almost no error during the operational period.
\Cref{fig:gpu_error_hardware} shows error propagation across GPU hardware components in A100 GPUs during the operational period.
We found three dominant GPU hardware error propagation paths, originating in (i)  GSP (GPU System Processor) errors, (ii) PMU (Performance Management Unit) SPI errors, and (iii) GPU Fallen Off the Bus errors.
We omit hardware error propagation graphs for H100 GPUs as we only observed three GSP errors and their propagation paths are similar to that of the A100 GPUs.

\textbf{GSP-related Errors.} 
Error propagations that originate in GSP-related errors are the most prominent among GPU hardware errors (see \Cref{tab:gpu_res_stats}) on A100 GPUs.
A GSP error arises when the GSP fails to respond to the remote procedure calls from the GPU driver. 
\Cref{fig:gpu_error_hardware} shows that, with a probability of 0.99, GSP errors lead to the recurrence of the same error or put the GPU in an inoperable state.
The remaining 0.01 (15 cases) of GSP errors caused PMU SPI communication errors (see the follow-up description) that led to user job failure, as depicted in~\Cref{sec:app}. 
Our analysis additionally shows that 99\% of GSP errors appeared in isolation without a preceding error.

GSP errors can be caused by either GSP firmware bugs~\cite{nv_driver_release_notes} or demanding workload. 
For example, \system SREs observed that these errors were highly correlated with heavy ML benchmarks, and they suggested that GSP errors are high-impact errors whose recovery requires manual node draining and reboots. 
Our propagation analysis confirm that the GSP is a single point of failure on both A100 GPUs in part because of their spontaneous nature and high downstream impact (e.g., GPU hangs) on the GPU.

\textbf{PMU SPI Errors.} Communication errors with the performance management unit (PMU) over the Serial Peripheral Interface (SPI), known as \emph{PMU SPI errors}, can cause performance management issues (e.g., inability to change the core frequency). 
We observed that such errors could lead to MMU errors with a probability of 0.88 (see~\Cref{fig:gpu_error_hardware}), ultimately leading to user job failures. 
The majority of the rest of the PMU SPI errors resulted in another PMU SPI error in close succession, leading to persisting error patterns. 
We observed 77 occurrences of PMU SPI errors (see~\Cref{tab:gpu_res_stats}) with a 0.98 probability of leading to user job failures (see~\Cref{tab:gpu_xid_app_fail}) on A100s.

\textbf{GPU Fallen Off the Bus.} GPU Fallen Off the Bus errors were logged when the GPU driver could not reach the GPU over the system bus.
This error is an integration error often caused by a loose GPU-motherboard connection or contact failure because of thermal cycles~\cite{tiwari2015reliability}.
Over 99\% of the errors of this type lead to similar errors in close successions and eventually put the GPU into an error state. 

Our \textit{GPU hardware error propagation} analysis suggests that the error detection and recovery of GSP, PMU, and the communication interfaces (e.g., PMU SPI) need to be improved via duplications and error-detection and correction mechanisms to prevent single points of failure, as evident in A100 GPUs.
In fact, AWS recommends disabling GSP for stability over performance benefits~\cite{aws_trouble}. 
By improving GSP hardware and driver software combined with a tightly integrating CPU-GPU architecture, the H100 GPUs in the GH200 Superchips significantly improve GPU hardware resilience over the A100 GPUs. 
Notably, apart from three GSP errors, which follow the same propagation paths as GSP errors in A100 GPUs,
we observed no other hardware error propagation paths on H100 GPUs.

\subsubsection{\textbf{NVLink Interconnect-Error Propagation.}}

\begin{figure}[t!]
    \centering
    \includegraphics[width=0.9\linewidth]{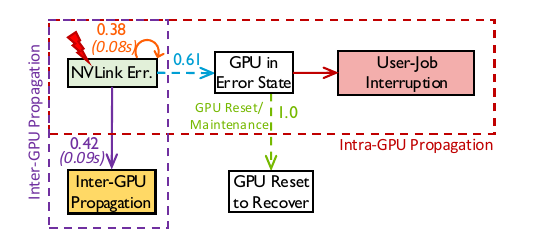}
    \caption{\label{fig:gpu_nvl_prop}NVLink intra-GPU and inter-GPU error propagation in A100 GPUs. Numbers on the edges show propagation probabilities and average propagation time in seconds.
    }
    \Description{Graph showing NVLink error propagation probabilities.}
\end{figure}

NVLink is an GPU-to-GPU interconnect for communication and data exchanges.
An NVLink error can impact a single or multiple GPUs on the same node, possibly rendering the entire multi-GPU compute pool unavailable (see \Cref{fig:gpu_nvl_prop}).
We observed both kinds of propagation in our error logs on A100 GPUs during the operational period on A100 GPUs.

\textbf{NVLink Inter- and Intra-GPU Propagation.} 
An NVLink error occurs when one or more NVLinks experience an error. 
Our analysis showed that of the 1,922 NVLink errors, 42\% propagated to connected GPUs; of those errors, $17\%$ involved three or more GPUs of the same compute node.
The rest of the 1,121 NVLink errors did not propagate across GPUs.
NVLink errors, like GPU hardware errors, happen in isolation without preceding errors. 
On the same GPU device, an NVLink error either leads to another NVLink error soon after (with a probability of $0.38$) or potentially leaves the faulty GPU in an error state (with a probability of $0.61$).

Although \system SREs reported that most NVLink errors were benign, we found that the probability of leading to user job failure when encountered is 54\% (43 cases) during the operational period on A100 GPUs. 
Moreover, in two incidents, GPU resets are needed to recover from critical NVLink errors, leading to over 2 hours of node interruption, and, as with GPU hardware errors, we found no preceding hardware errors before NVLink errors, making them less predictable than memory-related errors.

We did not observe NVLink-related errors on H100 GPUs during the operational period despite observing related non-error NVLink events (e.g., link initialization). 
We additionally conducted NVLink tests on an~\system-AI node to confirm that NVLinks are enabled and fully functional.
We conjecture that the improvement is due to both NVLink hardware and driver upgrades and potential changes in NVLink error logging mechanisms.

\begin{figure}[!t]
    \centering
    \includegraphics[width=0.99\linewidth]{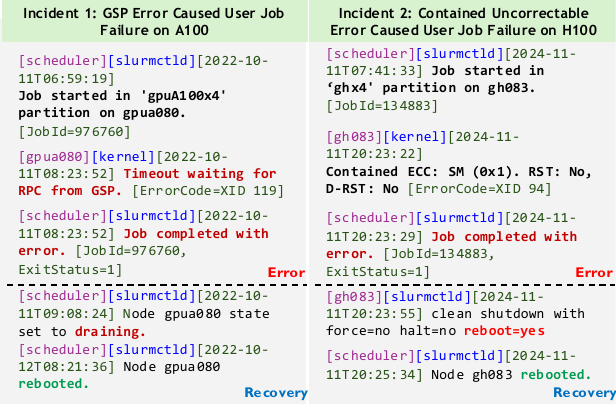}
    \caption{\label{fig:app_failed}Incidents in which GPU errors led to job failure.}
\Description{Code snippet showing GPU error incidents leading to job failure.}
\end{figure}

\begin{figure}[b]
    \centering
    \subfloat[GSP Error\label{fig:gsp_gpu_utilization}]{
        \includegraphics[width=0.49\linewidth]{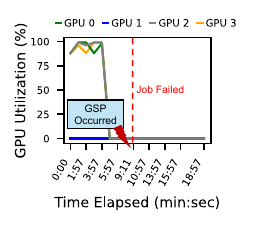}
    }
    \subfloat[Uncorr. Mem. Error\label{fig:dbe_gpu_utilization}]{
        \includegraphics[width=0.49\linewidth]{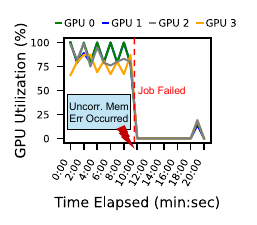}
    }
    \caption{GPU utilization during GPU error incidents.}
    \label{fig:gsp_dbe_gpu_utilization}
\end{figure}

\subsection{Examples of Error Propagation to User Jobs}

In this section, we show how measured errors relate to NVIDIA's measured DCGM GPU utilization metric, a step towards relating GPU memory and hardware errors with their impact on the user jobs.
\Cref{fig:app_failed} presents two example incidents.

\textbf{Incident 1:}
A GSP error stalled GPU control functions and rendered the GPU inoperable on an A100 GPU.
Consequently, the user job scheduled on that GPU failed. 
The GSP error required the draining of the node and a full node reboot to recover, which led to the draining of all pending user jobs on that node.
From the beginning of the node drain to the completion of the node reboot, the total recovery time for this incident was 23 node hours (09:08 AM to 08:21 AM the next day), during which the node was unavailable.
Figure \ref{fig:gsp_gpu_utilization} shows that the corresponding GPU utilization dropped quickly following the error incident due to job failure.
This incident shows that a GPU error can significantly interrupt user jobs and node availability.

\textbf{Incident 2:}
In this incident, the H100 GPU running the user job experienced an uncorrectable ECC memory error, which led to an error containment event. 
The error containment event contains the uncorrectable memory error by terminating the user's job.
Subsequently, a node reboot is triggered to recover from this error, resulting in a two minute node downtime.
Figure \ref{fig:dbe_gpu_utilization} shows that GPU utilization dropped as the user job was terminated.
This incident shows that although an uncorrectable memory error might be contained, it still results in user job termination and requires node reboot to fully recover, resulting in unexpected node downtime.
\section{Propagation of Errors to Jobs}
\label{sec:app}
This section provides an in-depth analysis of job-level resilience and associated GPU downtime. 

A GPU error can lead to:

(i) \emph{Job Failure:} A GPU error may not be handled by jobs, either because the error itself is not contained or because there is a lack of appropriate error-handling mechanisms. For example, GSP errors lead to GPU failures and require node reboots. To recover from GPU failures, jobs need to be re-executed from the beginning or rolled back to the closest checkpoint.

(ii) \emph{GPU and Node Downtime:} Downtime can occur when GPUs need to be reset or replaced by the operator, and no jobs can be scheduled on the GPU and the corresponding node in the interim.

\subsection{Result Highlights}

\Cref{tab:gpu_xid_app_fail} show the overall impact of hardware errors on applications. In summary, we observe that:

(i) Except for MMU and NVLink errors, no other GPU errors are handled by jobs, thus resulting in their failure.
Depending on the GPU type, the underlying cause of job failures differs.
In the case of A100 GPUs, hardware errors predominantly lead to job failures, whereas memory errors are the primary cause of H100 GPUs.
Therefore, there is a need to improve the resilience of underlying hardware to minimize such failures.
While checkpointing is an option, checkpointing routines have a high overhead of up to 40\%~\cite{wang2005modeling,wan2024bytecheckpoint,maurya2024datastates}, including management, storage, and restore.

(ii) The overall availability per-GPU node is $\sim$99.4\% (corresponding to 9 minutes downtime/day) and $\sim$99.3\% (corresponding to 10 minutes downtime/day) for A100 and H100 GPUs respectively.
This level of unavailability suggests that even if the rest of the infrastructure is highly available, current GPUs may not provide sufficient availability to meet the demands of critical applications that require greater than 3 9's of availability (downtime of 1.4 mins/day).
In addition, we used emulation to project the impact of this availability distribution at increased scales (for both node scale and job duration) and found that significant overprovisioning of  5\% would be necessary to handle associated failures (as explained in further detail in~\Cref{sec:gpu_dt}).

\begin{figure}[t]
    \centering
    \subfloat[A100\label{fig:gpu_reschedule_left}]{
        \includegraphics[width=0.49\linewidth]{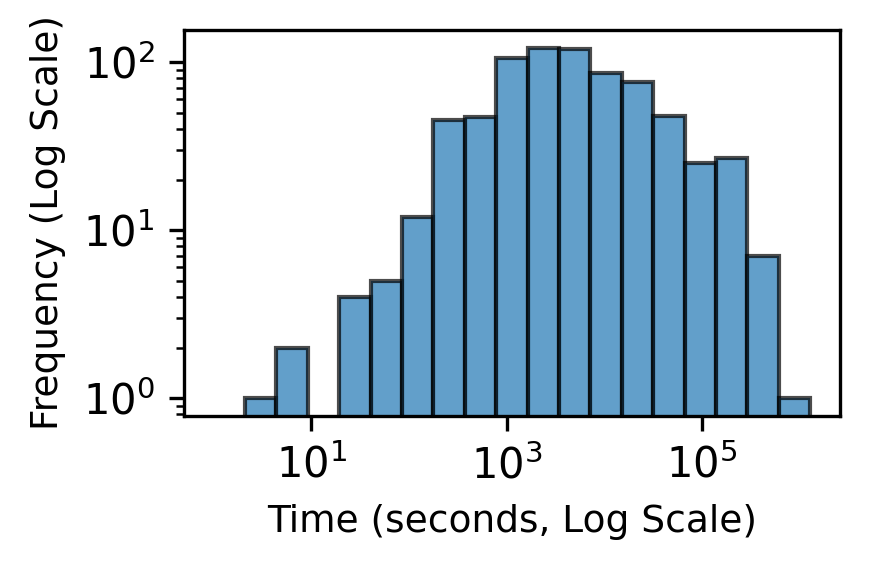}
    }    \subfloat[H100\label{fig:gpu_reschedule_right}]{
        \includegraphics[width=0.49\linewidth]{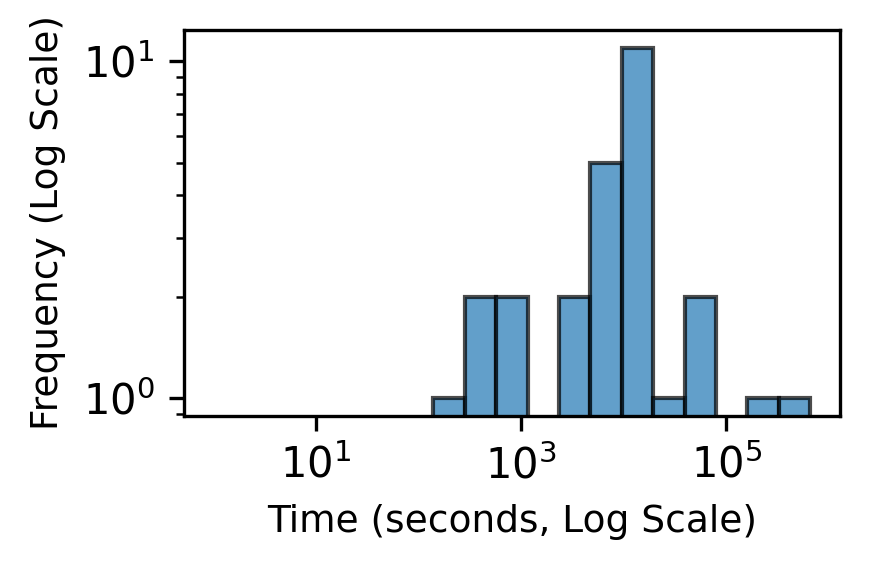}
    }
    \caption{Node unavailability time after GPU failures.}
    \label{fig:gpu_reschedule_combined}
\end{figure}

\subsection{Job Statistics}

During the characterization period, 1,420,278 user jobs were submitted to GPU nodes with a success rate of 87\%.
About 74\% of user jobs ran on a single GPU; 24\% ran on 2--4 GPUs; and only 2\% of jobs used five or more GPUs. 

Because of a lack of specific information on whether jobs were ML-related, we estimated the percentage of ML user jobs based on job submission names and the system modules/libraries imported\footnote{Due to privacy restrictions, we could not access the job scripts for use in job classification.}.
For instance, user jobs with names containing \verb|model| or \verb|train| were likely related to machine learning. 
We provide detailed statistics on node hours used, durations for both job types, and failure probabilities in \Cref{tab:distribution_job_gpu_hours}.
We find that across all job sizes, the failure probability lies between 6--49\%, indicating a lack of adequate recovery mechanisms.

\subsection{Job Failure Analysis}

To understand job failure patterns, we first separated jobs into ``Completed'' and ``GPU-Failed'' depending on their completion status. 
Based on that job categorization, we analyzed (i) the GPU errors that were most likely lead to a job failure, and (ii) the potential recovery strategies, such as checkpointing and exception handling.

\noindent \emph{Classifying Job Runs: }
We classified jobs based on their exit status and proximity of job failure time to GPU error occurrence time.
The job exit status was obtained from the Slurm job scheduler logs (as described in~\Cref{sec:data}).
We marked a job as ``GPU-Failed'' if a GPU error occurred within a 20-second interval before job failure.\footnote{Note that we do not count ``zombie'' jobs, i.e. jobs that have failed but not terminated by the Slurm scheduler in the analysis.}

\begin{table}[b!]
\centering
\resizebox{0.95\linewidth}{!}{
\begin{tabular}{c|c|cc|cc|cc}
\toprule
\textbf{XID} & \textbf{
\makebox[0pt][c]{\makecell[c]{GPU \\ Error}}} &
\multicolumn{2}{c|}{\textbf{\makecell[c]{\# GPU-failed \\ Jobs}}} &
\multicolumn{2}{c|}{\textbf{\makecell[c]{\# Jobs \\ Encountering \\ Given XID}}} &
\multicolumn{2}{c}{\textbf{\makecell[c]{Job Failure \\ Probability (\%)}}} \\
& & \textbf{A100} & \textbf{H100} & \textbf{A100} & \textbf{H100} & \textbf{A100} & \textbf{H100} \\
\midrule
31  & MMU err.               & 3206 & 93  & 3543 & 126 & 90.48 & 73.80 \\ \hline
74  & NVL err.               & 43   & 0   & 80   & 0   & 53.75 & --  \\ \hline
122 & SPI PMU RPC failure    & 40   & 0   & 41   & 0   & 97.56 & --  \\ \hline
119 & GSP RPC timeout        & 31   & 0   & 31   & 0   & 100.00 & -- \\ \hline
94  & Contained ECC          & 5    & 5   & 5    & 5   & 100.00 & 100.00 \\ \hline
48  & GPU DBE                & 0    & 5   & 0    & 5   & --  & 100.00 \\ \hline
64  & Row remapping failed   & 0    & 3   & 0    & 3   & --  & 100.00 \\ \hline
63  & Row remapping event    & 0    & 2   & 0    & 2   & --  & 100.00 \\ 
\bottomrule
\end{tabular}
}
\caption{\footnotesize Distribution of GPU-failed jobs across the different GPU error types for A100 and H100 GPUs. The failure probability is calculated as (\# GPU-failed jobs encountering that GPU error) / (\# jobs encountering that GPU error). The total number of GPU-failed jobs was 3,359 during the 895-day characterization period.}
\label{tab:gpu_xid_app_fail}
\end{table}

\noindent \emph{Correlating GPU Errors and Job Failures: }
We broke down all GPU-Failed jobs by the specific GPU errors that were most likely to lead to job failures.
Table~\ref{tab:gpu_xid_app_fail} provides probabilities of user job failures per GPU error for the A100 and H100 GPUs.
As any of the encountered errors may have contributed to a job failure, we consider all GPU errors that occurred within the 20-second interval to be responsible for the failure.

Overall, other than NVLink and MMU errors, all GPU errors, such as GSP RPC timeout and PMU failures, propagate (as discussed in~\Cref{sec:error_prop}) and cause job failures.
Based on previous analysis, we note that hardware errors -- such as GSP and PMU errors are dominant in A100 GPUs, whereas memory errors are dominant in H100 GPUs.

NVLink and MMU errors do not necessarily lead to job failures because: (1) For \textit{NVLink errors}, the link or GPU may not be in use by any user jobs (as discussed in~\Cref{sec:error_prop})\footnote{Based on our understanding, NVLink and memory errors can occur even when no workload is running~\cite{nvidia_forum_issue}.}; and (2) For \textit{MMU errors}, there can be application or library-level masking mechanisms. Besides hardware errors, MMU errors can also occur if buggy user code makes illegal memory accesses that cannot be mapped in the virtual-to-physical address space.
    Such errors can be managed using appropriate application-level exception handlers.
    Popular libraries and frameworks for machine learning~\cite{pytorch_distributed_exception_handling,pytorch_issue_1137,pytorch_lightning_discussion_15188} have support for handling such exceptions by skipping the associated training iteration, albeit at the cost of model quality.

\subsection{Impact of GPU Downtime on Jobs}
\label{sec:gpu_dt}

\renewcommand{\arraystretch}{1.3}
\setlength{\tabcolsep}{5pt}

\begin{table}[!t]
\centering
\resizebox{0.99\columnwidth}{!}{%
\begin{tabular}{l|l|l|l|l|ll}
\toprule
\textbf{GPU}         & \textbf{Count (\%)}  & \multicolumn{2}{c|}{\textbf{Elapsed Time (Minutes)}}           & \textbf{Failed (\%)}           & \multicolumn{2}{c}{\textbf{GPU Hours (k)}}  \\ \cline{3-7} 
\textbf{Count}       &                      & \textbf{Mean} & \textbf{P99} &                                 & \textbf{ML} & \textbf{Non-ML} \\ 
\toprule
1         & 1,052,993 (74.140\%)     & 134.560 & 2859.677  & 129,395 (12.29\%) & 641.9  & 1,949.14 \\
2-4       & 337,637 (23.773\%)       & 159.839 & 2880.117  & 40,610 (12.03\%)  & 810.4  & 2,749.00 \\
5-8       & 13,907 (0.979\%)         & 231.927 & 2880.316  & 4,469 (32.13\%)   & 253.1  & 290.16  \\
9-32      & 13,378 (0.942\%)         & 221.636 & 2880.167  & 3,045 (22.76\%)   & 307.5  & 844.40  \\
33-64     & 1,375 (0.097\%)          & 145.206 & 2880.017  & 608 (44.22\%)     & 108.2  & 149.24  \\
65-128    & 856 (0.060\%)            & 320.185 & 2834.413  & 421 (49.18\%)     & 15.1   & 442.70  \\
129-256   & 110 (0.008\%)            & 174.713 & 2041.312  & 7 (6.36\%)        & 0.0    & 57.23   \\
257+      & 22 (0.002\%)             & 29.128  & 107.493   & 5 (22.73\%)       & 0.0    & 3.44    \\
\bottomrule
\end{tabular}%
}
\caption{\footnotesize Job distribution, elapsed time statistics (mean, P99), failure count with percentage, and GPU hours divided into ML and non-ML categories for various A100 and H100 GPU configurations.}
\label{tab:distribution_job_gpu_hours}
\end{table}

While significant node hours might be lost because of wasted compute time from failed jobs, additional node hours are also lost because of the time required to recover the impacted GPU node by either resetting it or replacing it entirely. 
To reset the GPU node, operators typically drain the node, i.e., wait for other jobs running on the node to complete without accepting new jobs and then reboot.
After the reboot, if the node successfully passes the health check, the node reset is successful, and new jobs can be scheduled on the GPU node.
If the reset is unsuccessful, the node is marked failed until the GPU is additionally tested and physically replaced, if required.
To calculate the average system downtime, we estimated the total time when the GPU was unavailable, which primarily included the drain and reboot time.
\Cref{fig:gpu_reschedule_combined} shows the distribution of the unavailable time across the entire characterization duration. 
Overall, we found that the expected time to service the failed node was 0.88 hours for A100 GPUs and 2.2 hours for H100 GPUs.
A total of 5,700 node hours were lost to GPU downtime.
Using the node downtime and failure distributions, we can estimate the availability of the GPU node as $\frac{MTTF}{MTTF+MTTR}$ equal to 99.4\% and 99.3\% for A100 and H100 GPUs respectively~\footnote{The node MTTF number is estimated from the GPU's MTBE, for which we conservatively assume that all critical GPU errors lead to node interruption.}.

\noindent \emph{Projected impact of availability on long-running and large-scale jobs: }
We provide error and recovery statistics in previous sections; we also attempted to project how those distributions would affect jobs running on a different system.
To do so, we built a simulation tool driven by our analysis.
The parameters of the simulation tool can be varied based on the scenario under consideration.

Specifically, we simulate the case in which jobs (such as ML training) use the entire set of 608 H100 GPUs and run for a duration of 1 month.
These jobs require all GPUs to be operational to make progress, and frequent node failures can lead to resource unavailability and slower job progress.
When such failures occur, additional provisioning of GPU resources is necessary to allow the job to resume on alternate nodes while the failed nodes recover.

The simulation uses a discrete time event simulation with node failure probabilities derived from our prior analysis.
The recovery time after a failure is dependent on variables such as checkpoint load time and availability of spare GPUs.
To account for the variability introduced by these factors, we parameterize recovery time and perform a parameter sweep.
For a training job with 608 GPUs and recovery time of 2.2 hours, the required overprovisioning is 5\%: i.e., 31 additional GPUs are needed beyond the original 608 to maintain availability of 99.9\% at the job level.
While at first glance, such overprovisioning would appear to be a small cost, for the above example, it would cost over \$1 million per month for a 1000 node cluster (our analysis is based on AWS H100 GPU rental rates~\cite{gpu_costs}).
However, if the recovery time is reduced to 5 minutes, downtime decreases significantly, and the required overprovisioning drops to 2\%, a $2.5\times$ reduction.
This highlights the criticality of minimizing recovery time to reduce downtime for large and long-running jobs.

In summary, every GPU node in the system has ``two nines'' of availability.
While such availability does not significantly impact small jobs that use recovery mechanisms, large jobs can face significant downtime.
Significant overprovisioning of up to 5\% is required to eliminate such downtime.

\section{Discussion}
\label{s:discussion}

\textbf{Justification of Analyzing Errors.}
Data-driven HPC resilience characterization studies analyze operational data on system/application errors to provide insights into system resilience~\cite{debardeleben2014gpu,di2014lessons,tiwari2015understanding,tiwari2015reliability,gupta2015understanding,gupta2017failures,ostrouchov2020gpu,nie2016large,nie2017characterizing,oles2024understanding}. 
While in those studies, the error rate is used as the key metric to quantify resilience, some~\cite{vilas_2015,Beigi2023} argue that fault rate is a more appropriate metric. 
Errors represent the manifestation of faults and have direct downstream consequences, such as triggering of recovery mechanisms, application interruptions, or system-wide outages (SWOs). 
While a fault may result in multiple errors, it is the resulting errors that the recovery mechanisms must address to maintain system health. 
These errors and their recovery process directly impact system health, performance, and availability.
Thus, SREs prioritize errors over faults.
Hence, like many others who study operational data, we chose to study errors.

\textbf{Reliability of Logging.}
A potential source of error is logging inconsistency from (i) missing logs due to storage failures, (ii) node lockups where the NVIDIA driver fails to log the error prior to the failure, and (iii) incorrect job-status captured by Slurm. However, these issues negligibly affected our analysis because: 
(1) \system minimizes storage failure impact by streaming logs in real time to centralized storage; 
(2) SRE records show only 27 A100 and nine H100 node lockups--0.18\% and 0.49\% of total GPU errors--making their impact negligible; and 
(3) while eliminating false-negative job status is challenging, GPU errors studied led to job failure almost 100\% of times (\Cref{tab:gpu_xid_app_fail}), indicating failure statuses were reliably captured by Slurm and false-negatives minimally impacted our findings.

\section{Related Work}
Existing work has analyzed GPU resilience at the microarchitecture, cluster, and application levels. This paper extends existing work via comprehensive analyses of GPU error characteristics, propagation, and impact on user jobs.

\textit{Microarchitecture-level GPU Resilience}. Previous research~\cite{vallero2018multi,sartzetakis2022gpufi,hari2015sassifi,yang2024gpu} has primarily focused on the resilience of individual GPUs at the microarchitecture and software levels, e.g., for older generations such as the NVIDIA G80~\cite{braga2023software}. However, the earlier work did not evaluate the resiliency of modern GPUs in large-scale HPC settings.

\textit{System-level GPU Resilience in HPC Settings}.
Existing studies have analyzed the resilience of GPUs in HPC systems~\cite{haque2010hard,debardeleben2014gpu,kokolis2024revisiting,gunawi2018fail,10.1145/3372790_comment}, for example, the NVIDIA Tesla K20X GPUs in various supercomputers \cite{di2014lessons,tiwari2015understanding, tiwari2015reliability,gupta2015understanding,nie2016large, gupta2017failures,nie2017characterizing,ostrouchov2020gpu}. Studies of the Blue Waters, Titan, and Summit supercomputers~\cite{di2014lessons,nie2016large,nie2017characterizing,oles2024understanding,kokolis2024revisiting} have examined node failures and GPU error characteristics. 
However, such work either studied previous generations of GPUs with a focus on GPU memory or on cluster-level resilience instead of GPU resilience. Our work complements those by providing resilience insights into the latest generation of GPUs, focusing on a broader range of components.

\textit{Application-level GPU Resilience in Data-centers and Deep Learning Workloads.}
Recent research has focused on understanding GPU power usage~\cite{nie2017characterizing}, GPU component-level failures~\cite{jung2023predicting}, software-level error handling~\cite{maruyama2010high,fang2014gpu,lian2025}, and the impact of GPU software error propagation on GPGPU applications~\cite{wei2020g,li2016understanding,anwer2020gpu,rech2014impact} and emerging GPU workloads such as convolutional neural networks (CNNs)~\cite{dos2018analyzing,condia2022multi}, large language models (LLMs)~\cite{jiang2024megascale,dubey2024llama}, and privacy/safety-critical applications~\cite{li2024advances,perez2022gpu,shah2024characterizing}.

\section{Conclusion}
This paper describes the results of a resilience study of \system, which consists of 1,056 NVIDIA A100 and H100 GPUs. 
The study used up to 2.5 years of operational data on GPU errors collected across those GPUs. 
We assessed the resilience of GPU components, error propagation paths, and impact on jobs and compared the two generations of GPUs. 
In future work, we will extend the analysis presented to other accelerators and larger-scale systems, running more complex HPC and ML workloads.

\begin{acks}
We thank the reviewers for valuable feedback and J. Applequist, H. C. Fairow, S. Weick, and K. Atchley for support. This work used NCSA \system and \systemai compute resources supported by NSF grants 2005572 and 2320345, with allocations from ACCESS (NSF 2138259, 2138286, 2138307, 2137603, 2138296), and additional support from NSF grants 2029049, 2319190, 2430244, 2530738, the IBM-Illinois Discovery Accelerator Institute, and the VinUni-Illinois Smart Health Center. Any opinions expressed here are those of the authors and do not reflect the views of NSF, IBM, or VinUni.
\end{acks}

\bibliographystyle{ACM-Reference-Format}
\bibliography{bibliography}

\end{document}